\documentclass[a4paper,11pt]{article}

\usepackage{jheppub} 

\usepackage[T1]{fontenc} 

\usepackage{amsmath,amssymb,amsfonts}
\usepackage{slashed}
\usepackage{graphicx}
\usepackage{float} 
\usepackage{subfigure}

\usepackage{verbatim} 
\usepackage{ulem} 
\usepackage{latexsym}
\usepackage{color}
\usepackage{bm}

\def\be{\begin{eqnarray}}
\def\ee{\end{eqnarray}}

\title{\boldmath Baryons as Vortexes on the $\eta^{\prime}$ Domain Wall}

\author[a,b,c]{Fan Lin}
\author[d,a,e,1]{Yong-Liang Ma,\note{Corresponding author.}}

\affiliation[a]{School of Fundamental Physics and Mathematical Sciences, Hangzhou Institute for Advanced Study, UCAS, Hangzhou, 310024, China}
\affiliation[b]{Institute of Theoretical Physics, Chinese Academy of Sciences, Beijing 100190, China}
\affiliation[c]{University of Chinese Academy of Sciences, Beijing 100049, China}
\affiliation[d]{School of Frontier Sciences, Nanjing University, Nanjing University, Suzhou, 215163, China}
\affiliation[e]{International Center for Theoretical Physics Asia-Pacific (ICTP-AP) , UCAS, Beijing, 100190, China}

\emailAdd{linfan19@mails.ucas.ac.cn}
\emailAdd{ylma@nju.edu.cn}

\abstract{We show that the recent construction of $N_f=1$ baryons on the $\eta^\prime$ domain wall can be understood as vortexes of the principal effective theory---the Chern-Simons-Higgs theory---on a 2+1-dimensional sheet. This theory has a series of vertex solutions, and the vortex with unit topological charge naturally spins $N_c/2$, which coincides with the spin of the one-flavor baryon in QCD. Since the $N_c$ scaling of the vortexes is the same as that of baryons, baryons can be regarded as vortexes. By virtue of the particle-vortex symmetry, the dual Zhang-Hansson-Kivelson theory indicates that the quark carries topological charge $1/N_c$ and obeys fractional statistics. The generalization to arbitrary $N_f$ is also discussed.}

\begin{document}

\maketitle
\flushbottom

\allowdisplaybreaks{}

\section{Introduction}

At the low-energy region, Quantum Chromodynamics (QCD) becomes a strongly coupled system, and, quarks and gluons are confined to colorless mesons and baryons. Among these mesons, the lightest pseudoscalar mesons which can be identified with the Nambu-Goldstone bosons generated by chiral symmetry breaking are particularly interesting, as they can be well described by the nonlinear chiral dynamics and carry intrinsic topology of QCD. In the large $N_c$ limit, QCD is extremely simplified and dominated by planar diagrams~\cite{tHooft:1973alw}. In this limit, baryons can be regarded as solitons made up of interacting mesons since their properties exhibit similar $N_c$ scalings~\cite{Witten:1979kh}.

Chiral solitons consisting of nonlinearly interacting pion fields, known as skyrmions, were suggested to be baryons before the establishment of QCD~\cite{Skyrme:1961vq,Skyrme:1962vh}. These chiral solitons in 3+1 dimensions are protected by a non-trivial homotopy group $\pi_3(SU(N_f))=\mathbb{Z}, N_f \geq 2$ with integer $\mathbb{Z}$ being the winding number. The conserved winding number due to topology can be taken as the conserved baryon number in QCD. However, for the one-flavor case, there are no topology-protected soliton solutions to describe baryons due to the trivial homotopy group $\pi_3(U(1))=0$. This was regarded as the fatal drawback of the skyrmion approach to baryon physics.

Recently, it has been argued that, under the large $N_c$ limit, it is possible to construct one-flavor baryons as Hall droplets on $\eta^\prime$ domain walls~\cite{Gaiotto:2017yup,Gaiotto:2017tne}. The  $\eta^\prime$ particle field exhibits $2\pi$ periodicity and has a series of equivalent ground states $\eta^\prime=0\;\mathrm{mod}\;  2\pi$. Therefore, domain wall structures exist in four-dimensional spacetime. The domain wall is a 2+1 dimensional sheet defined at $\eta^\prime=\pi$. After imposing appropriate boundary conditions, the sheet behaves like a baryon for one-flavor and, also for arbitrary flavors after avoiding the Nambu-Goldstone bosons on the wall. When the domain walls are bounded by axionic strings, the axionic domain walls can carry a baryonic charge representing the low energy description of the baryons made by the extra quark flavor~\cite{Bigazzi:2022ylj}.

In fact, the $\eta^\prime$ domain wall supports a topological $SU(N_c)_{-1}$ Chern-Simons (CS) theory concerning the gluon field~\cite{Komargodski:2018odf}. By level-rank duality, it is more convenient to work with the abelian $U(1)_{N_c}$ CS theory for the purpose to understand how to couple the baryon background gauge field to the $SU(N_c)_{-1}$ CS theory. Typically, the $U(1)_{N_c}$ CS theory describes the fractional quantum Hall effect with fractional filling number $1/N_c$. Thus, the sheet with a boundary is precisely a quantum Hall droplet, where the corresponding edge mode carries topological charges, namely baryon number.

Baryons as quantum Hall droplets also can be understood as chiral bags in a (2+1)-dimensional strip using the Cheshire Cat principle \cite{Ma:2019xtx}. For a small bag radius, the bag reduces to a vortex line which is the smile of the cat with flowing gapless quarks all spinning in the same direction. The disk enclosed by the smile is described by an emergent topological field theory due to the Callan-Harvey anomaly outflow~\cite{Callan:1984sa}. The chiral bag naturally carries the unit baryon number and spin half $N_c$. Besides, based on the Witten-Sakai-Sugimoto model, Hall droplet sheets in holographic QCD were proposed recently~\cite{Bigazzi:2022luo}.

When the chiral effective theory of pseudoscalar mesons is extended to include vector mesons through the hidden local symmetry (HLS) approach, in addition to the intrinsic-parity even terms, there is an intrinsic-parity odd sector, the homogeneous Wess-Zumino terms~\cite{Bando:1984ej,Bando:1987br,Harada:2003jx}. Some homogeneous Wess-Zumino terms introduce the coupling between iso-scalar vector meson $\omega$ and winding number current---baryon number current. This means that $\omega$ meson field acts as the source of baryon current. Considering that these homogeneous Wess-Zumino terms couple to $\eta^\prime$ field on the domain wall, a $U(1)_{N_c}$ Chern-Simons (CS) theory about vector meson $\omega$ field which can be identified as the emergent gauge field emerges naturally~\cite{Karasik:2020pwu,Karasik:2020zyo}. The level-rank duality essentially describes the duality between $SU(N_c)_{-1}$ gluon field and $U(1)_{N_c}$ vector meson $\omega$ meson field.

All the investigations on the topological field theory only focus on the gluon side. To have a complete theory, it should be extended to include the strongly interacting fermions/quarks. In such a case, the dual theory involves more fields and the duality between these two theories is conjectured as~\cite{Gaiotto:2017tne,Hsin:2016blu}
\begin{equation}
	SU(N_c)_{-N_f} +N_f \;\mathrm{fermions} \longleftrightarrow U(N_f)_{N_{c}} + N_f\; \mathrm{scalars}.
	\label{dualityNf}
\end{equation}
So that the dual theory on the right involves $N_f$ scalar fields that correspond to the bosonization of the fermions/quarks on the left-hand side. The duality~\eqref{dualityNf} sets up a relation between the color gauge symmetry associated with the $SU(N_c)$ group and the flavor gauge symmetry associated with the $U(N_f)$ group. In other words, the global flavor symmetry is gauged under level-rank duality.

Duality~\eqref{dualityNf} tells us that, in the case of one flavor, care must be taken to investigate the scalar and topological CS fields together. Actually, it is found that the scalar is crucial for a concrete description of the baryon on the $\eta^\prime$ domain wall. With gauge symmetry assumed, the effective theory on the domain wall is shown to involve a complex scalar in Higgs phase that minimally couples with a $U(1)_{N_c}$ CS theory --- the Chern-Simons-Higgs theory. This theory possesses several vortex solutions, each of which carries a topological charge. Their qualities, such as mass and radius, exhibit similar behaviors to that of one-flavor baryons in the large $N_c$ limit. By the particle-vortex symmetry, a dual description is presented, which turns out to be the Zhang-Hansson-Kivelson theory \cite{Zhang:1988wy}, originally proposed for studying electrons in the context of the fractional quantum Hall effect. This theory is applied on the domain wall to quarks, showing that quarks carry a $1/N_c$ baryon number and obey fractional statistics, has been pointed in \cite{Ma:2019xtx}.

\section{Baryons as vortexes for $N_f=1$}

In the large $N_c$ limit, the $\eta'$ meson becomes a massless Nambu-Goldstone boson associated with the breaking of the $U(1)_A$ symmetry~\cite{Witten:1979vv}~\footnote{Throughout this work, we will consider the chiral limit.}. The $\eta'$ field is a periodic field, parameterized on the circle $\eta'\simeq\eta'+2\pi$ and, in the effective theory of $\eta^\prime$ and, the $\eta'=0\;\mathrm{mod}\;  2\pi$ is the unique ground state of $\eta'$ field. When $1/N_c$ corrections are included, the domain wall $\eta'=\pi$ acts as a cusp for some heavy fields that need to be rearranged at this point. To remedy this, it turns out that the domain wall acquires a topological field theory, which is identified with an $SU(N_c)_{-1}$ Chern-Simons theory plus some matter fields.
 
For simplicity, we shall first focus on the one-flavor case with $N_f=1$ and then extend to arbitrary flavors later. When $N_f = 1$, the duality in Eq.~(\ref{dualityNf}) can be rephrased as
\be
SU(N_c)_{-1} +\mathrm{one \; fermion}\; \psi \longleftrightarrow U(1)_{N_{c}} + \mathrm{one\;scalar}\; \phi.
\label{duality1}
\ee
Therefore, the effective field theory of the right-hand side involves a $U(1)_{N_{c}}$ CS gauge field $a_\mu $ and a complex scalar field $\phi$. If we assume gauge invariance and minimal coupling, the following effective Lagrangian is conjectured:
 \be
 	\mathcal{L}_A[\phi,a] & = &  |\partial_\mu\phi-\mathrm{i}a_\mu\phi|^2+\frac{N_c}{4\pi}\epsilon^{\mu\nu\rho}a_\mu\partial_\nu a_\rho -V(\phi^*\phi),
 	\label{LA}
 \ee
where the scalar field $\phi$ is the bosonization of the fermion field $\psi$, that is, $\phi^*\phi\sim\psi^\dagger\psi$. The gauge field $a_\mu$ represents an emergent gauge field of the gauge group $U(1)$ associated with the global symmetry for baryon number conservation before it's gauged. This Lagrangian~\eqref{LA} emerges in the condensed physics where the fermion is an electron to depict the fractional quantum Hall effect with a $1/N_c$ filling fraction. Here, our fermion is quark, so the term $\phi^*\phi$ corresponds to the quark density and the emergent gauge field $a_\mu$ propagates quark number. Since quarks carry color charge, it's required that the density of quarks for every color, denoted as $\phi_{c}^*\phi_c=\phi^*\phi/N_c$, remains finite in the large $N_c$ limit. The potential $V(\phi^*\phi)$ is responsible for the non-zero vacuum expectation value $\langle\phi^*\phi\rangle=N_c  v^2$. Formally, it can be written as
\begin{equation}
	V(\phi^*\phi)=N_c\sum_{I=1}^{\infty} c_I (\frac{\phi^*\phi}{N_c}-v^2)^I,\;\;\; v>0,
\end{equation}
where the coefficients $c_I$ are subjected to the constraints to ensure that the non-zero vacuum expectation value of $\phi_c^*\phi_c$ equal to $v^2$ which is independent of $N_c$. 

It's worth mentioning that the Lagrangian $\mathcal{L}_A$~\eqref{LA} is of order $N_c$---the leading order of $N_c$ counting---with our choice of potential. Therefore, in the large $N_c$ limit, $\mathcal{L}_A$ dominates the physics on the domain wall. Other terms of order less than $N_c$ can also be included in the Lagrangian, but they have negligible effects in the large $N_c$ limit.

In model~\eqref{LA}, there are (2+1)-dimensional topological non-trivial finite-energy vortex configurations satisfying the equations of motion~\cite{Jatkar:1989sc}  
\begin{subequations}
\be
(\partial_\mu- i a_\mu)(\partial^\mu-i a^\mu)\phi +\frac{\partial V}{\partial \phi^*}=0, \\
(\partial_\mu+ i a_\mu)(\partial^\mu+ i a^\mu)\phi^* +\frac{\partial V}{\partial \phi}=0, \\
i(\phi^* \partial^\mu\phi-\phi\partial^\mu\phi^* )+2 a^0 \phi\phi^* +\frac{N_c}{2\pi}\epsilon^{\mu\nu\rho}\partial_\nu a_\rho=0.
\ee
\end{subequations}
In the following analysis of the vortex properties, we do not need to solve this complex set of ordinary differential equations therefore the explicit form of the potential function $V(\phi^*\phi)$ is not necessary. Without losing generality we consider a single vortex located at the origin. In polar coordinates, we can take the ansatz~\cite{Jatkar:1989sc}
\begin{equation}
	\phi(\mathbf{r})=\mathrm{e}^{in\theta}f(r),\; a_0(\mathbf{r})=A_0(r),\;\mathbf{a}(\mathbf{r})= \frac{A(r)}{r}(\sin\theta,-\cos\theta),
	\label{polar}
\end{equation}  
with the boundary conditions for the finite energy configuration
\be
& & f(\infty)=v,\;\; A_0(\infty)= 0,\;\; A(\infty) = n;\\
& & f(0)= 0,\;\; A_0(0)=c,\;\; A(0) = 0,
\label{BC}
\ee
where $c$ is a non-zero constant. We will see later that $n\in \mathbb{Z}$ labeled the winding number of the vortex solutions.

The Chern-Simons term in the Lagrangian density is topological and gives rise to a topological current
\begin{equation}
	j^\mu=\frac{N_c}{2\pi}\epsilon^{\mu\nu\rho}\partial_\nu a_\rho=\frac{N_c}{4\pi}\epsilon^{\mu\nu\rho} f_{\nu\rho},
	\label{eq:TopCharge1f}
\end{equation}
where $f_{\mu\nu}=\partial_\mu a_\nu - \partial_\nu a_\mu$ is the field strength tensor of $a_\mu$. From the topological current~\eqref{eq:TopCharge1f} one can see that the vortex solution carries topological charge
\begin{equation}
	Q=\int j^0 dxdy= \frac{N_c}{2\pi}\int \epsilon^{0\nu\rho}\partial_\nu a_\rho dxdy = nN_c,
\end{equation}
which is actually the quantization of vortex flux 
\begin{equation}
	\Phi=\int \epsilon^{0\nu\rho}\partial_\nu a_\rho dxdy=
	\int \mathbf{a}\cdot d\mathbf{r}=\int \frac{n}{r} rd\theta=2\pi n.
\end{equation}

As is known, objects that carry both flux and charge are anyons, which obey fractional statistics~\cite{Wilczek:1981du,Rao:1992aj}. The vortexes discussed above are anyons and have spins
\begin{equation}
	s=\frac{Q\Phi}{4\pi}=n^2\frac{N_c}{2},\;\; n\in \mathbb{Z}.
\end{equation}
It's straightforward to see that vortexes with $n = \pm 1$ have the same spin as one-flavor baryons in the ground state. Since the scalar field $\phi$ corresponds to the quark number and its coupling strength with $a_\mu$ is normalized to $1$, the topological charge $Q$ can be defined as quark number. Considering that a baryon consists of $N_c$ quarks, it's then natural to define the baryon number
\begin{equation}
B = \frac{Q}{N_c} = n.
\end{equation}
Can we consider vortexes with $n = \pm 1$ as (anti)baryons, and $|n|\ge 2$ as multi-baryon structures located on the domain wall? The answer is yes! These vortexes actually behave similarly to one-flavor baryons in the large $N_c$ limit. There is also a special solution with $n=0$, which means the vortex has a zero baryon number. Mentioning the relation on the quark density $\phi^*\phi\sim\psi^\dagger\psi$, we can predict this vortex depicts the quark condensate. Before going into more details, it's worth noting that vortexes with $|n|\ge 2$ can be thought of as multi-baryon structures on the domain wall and have similar counterparts in 3+1 dimensions, namely the Skyrmions with multiple baryon numbers.

In the large $N_c$ limit, we have demanded that the quark density of every color $\phi^{*}_{c}\phi_c= \phi^*\phi/N_c$ keeps finite. The Lagrangian~\eqref{LA} can be reexpressed as
\begin{equation}
	\mathcal{L}_A= N_c\left[|\partial_\mu\phi_c- ia_\mu\phi_c|^2-V_c(\phi_{c}^{*}\phi_c)+\frac{1}{4\pi}\epsilon^{\mu\nu\rho}a_\mu\partial_\nu a_\rho \right],
\end{equation}
where 
\begin{equation}
	V_c(\phi_c^*\phi_c)=V(\phi^*\phi)/N_c=\sum_{I=1}^{\infty} c_I (\frac{\phi^*\phi}{N_c}-v^2)^I=\sum_{I=1}^{\infty} c_I (\phi_c^*\phi_c-v^2)^I,\;\; v>0,
\end{equation}
which does not scale with $N_c$. Then, the $N_c$ appears only as a multiplier in the Lagrangian, and will not enter the equation of motion or vortex solution. Therefore, the sizes of the vortexes are only determined by the color-independent parameters. Then, following the reasoning in Ref.~\cite{Witten:1979kh}, one can argue that the radii of the vortexes are $\sim N_c^0 $, $\mathrm{energy (mass)} \sim N_c$, the vortex-vortex scattering amplitudes are of the order $N_c$ and the meson-vortex (baryon) scattering amplitudes are of order $N_c^0$. All these scalings are the same as the corresponding ones of baryons. Therefore, vortexes can be regarded as baryons on the $\eta^\prime$ domain wall.

\section{Particle-vortex duality for $N_f=1$}

In the last section, there are two types of particle excitations introduced. The first type is the quantized $\phi$ which represents quarks. The second type are vortexes, the soliton solutions carrying winding numbers. In 2+1 dimensions, it is possible to express the same physics using two different theories, the particle of one theory are related to the vortexes of the other, and vice versa. This is known as particle-vortex duality, which has been studied in the literature, for example Ref.~\cite{Peskin:1977kp}. By this duality, the theory $\mathcal{L}_A$ indeed has a dual description, famous as relativistic version of Zhang-Hansson-Kivelson (ZHK) theory \cite{Zhang:1988wy,Tong:2016kpv} 
\begin{equation}
	\mathcal{L}_B[\tilde{\phi},\tilde{a}]=  |\partial_\mu\tilde{\phi}-i\tilde{a}_\mu\tilde{\phi}|^2-\tilde{V}(\tilde{\phi}^*\tilde{\phi})+\frac{1}{4\pi N_c}\epsilon^{\mu\nu\rho}\tilde{a}_\mu\partial_\nu \tilde{a}_\rho,
	\label{LB}
\end{equation}
where $\tilde{\phi}$ is a complex scalar field, $\tilde{a}_\mu$ is a $U(1)$ gauge boson field and $\tilde{V}(\tilde{\phi}^\ast\tilde{\phi})$ is a Higgs-type potential. Based on the particle-vortex duality, we will claim that $\tilde{\phi}$ should be interpreted as baryons while vortexes in the $\mathcal{L}_B$ theory should correspond to quarks.

In theory $\mathcal{L}_B$, there are also a series of vortex solutions, but with a different topological current compared to those of $\mathcal{L}_A$ that we studied earlier
\begin{equation}
	\tilde{j}^\mu=\frac{1}{2\pi N_c}\epsilon^{\mu\nu\rho}\partial_\nu \tilde{a}_\rho.
\end{equation}
Then, one can obatain the topology charges of the vortexes in theory $\mathcal{L}_B$ as
\begin{equation}
	\tilde{Q}=\int \tilde{j}^0 dxdy= \frac{1}{2\pi N_c}\int \epsilon^{0\nu\rho}\partial_\nu \tilde{a}_\rho dxdy = \frac{n}{N_c}.
\end{equation}
by useing the same parameterizations~\eqref{polar} and boundary conditions~\eqref{BC}. Since the coupling between $\tilde{\phi}$ and $\tilde{a}_\mu$ in~\eqref{LB} is set to $1$ and $\tilde{\phi}^*\tilde{\phi}$ depicts baryon density, $\tilde{a}_\mu$ propagates a unit baryon number. We can therefore define the baryon number, which exactly equate to the topological charge $\tilde{Q}$. Additionally, the vortexes in the theory $\mathcal{L}_B$ not only possess fractional topological charge but also exhibit fractional statistical spin
\begin{equation}
	\tilde{s}=\frac{\tilde{Q}\Phi}{4\pi}=\frac{n^2}{2N_c},\;\; n\in \mathbb{Z}.
\end{equation}
Consider the basic vortex states with winding numbers of $\pm 1$, predicted to be (anti)quarks, which indeed have the same statistics as the quarks leaked from chiral bags~\cite{Ma:2019xtx}. Other vortexes with larger winding numbers are multi-quark structures. Especially, vortexes with winding number of ±$N_c$ carry a unit baryon number and spin of $N_c/2$, which exactly correspond to one-flavor baryons. For an observer living on the domain wall, the nature of particle statistics is colorful.

In QCD, baryons are composite particles made out of interacting quarks and gluons. In theory $\mathcal{L}_A$, the excitations of the $\phi$ field are advised as quarks, which are more basic than the baryons as vortexes that are induced. It is possible and natural to analyze baryon properties dependent on large $N_c$ even more insight on the mesons, have been presented in the last section. However, in the dual theory $\mathcal{L}_B$, baryons as excitations of $\tilde{\phi}$ are deemed to be basic, and behaviors dependent on a large $N_c$ are missing. We cannot determine more qualities except topological charges (baryon numbers) and spins.

\section{Baryons as vortexes for $N_f>1$} 

Next, we extend the above discussion on  one-flavor to the multi-flavor. In such a case, according to the level-rank duality Eq.~(\ref{dualityNf}), the global symmetry of flavors $U(N_f)$ is gauged. Therefore, the gauge field propagates quark number and isospin charge. If the minimal coupling is still valid, a Lagrangian similar to $\mathcal{L}_A$ can be written down
\begin{equation}
		\mathcal{L}_C[\bm{\phi},A]=  |\partial_\mu\bm{\phi}-\mathrm{i}A_\mu\bm{\phi}|^2-V_C(\bm{\phi}^\dagger\bm{\phi})+\frac{N_c}{4\pi}\epsilon^{\mu\nu\rho}\operatorname{Tr}\left( A_\mu\partial_\nu A_\rho-i\frac{2}{3} A_\mu A_\nu A_\rho\right),
	\label{LC}
\end{equation}
where $\bm{\phi}=(\phi^1,\phi^2,\dots,\phi^{N_f})^T$ are $N_f$ quark fields and the gauge fields $A_\mu$ take values in $\mathfrak{u}(N_f)$.

For simplicity and reality, we consider $N_f=2$ case and put $A_\mu$ in adjoint representation $A_\mu= A_\mu^a t^a,a=0,1,2,3$, and $t^a=(1/2,\bm{\sigma}/2)$ where $\bm{\sigma}$ is Pauli matrixes. The potential is generally written as:
\be
V_C(\bm{\phi}^\dagger\bm{\phi}) & = & N_c\sum_{I=1}^{\infty}c_{I} \left(\frac{\bm{\phi}^{\dagger}\bm{\phi}}{N_c}-v^2\right)^{I}.
\ee
with respect to the $U(2)$ invariance. We intuitively assume that vortex solutions still exist for this Lagrangian. Based on our previous observations, determining both the topological charge and statistical spin of the vortexes requires careful consideration of their asymptotic behavior. The potential determines the vacuum state for $\bm{\phi}$. For the sake of simplicity, we choose the vacuum state to be $\bm{\phi}_0=(v_1 e^{in_1 \theta},v_2 e^{in_2 \theta+i\varphi}), n_1,n_2\in \mathbb{Z}, v_1^2+v_2^2=v^2, v_1,v_2>0$ where $\varphi$ is an extra phase angle that cannot be fixed. Putting the center of the vortex at original point, one has $\bm{\phi}(r\rightarrow \infty, \theta)=\bm{\phi}_0$. To make the energy of vortexes finite, it is necessary
\begin{equation}
	\int \mathrm{d}\bm{r}^2|\partial_i\bm{\phi}-iA_i\bm{\phi}|^2 < +\infty. 
\end{equation}

By only considering terms up to $o(\frac{1}{r})$, we can impose constraints on the gauge field that describe its behavior at infinity as
\begin{equation}
	 A_i^0(r\rightarrow \infty, \theta)=(n_1+n_2)\frac{\mathbf{e}_\theta}{r}, \quad A_i^3(r\rightarrow \infty, \theta)=(n_1-n_2)\frac{\mathbf{e}_\theta}{r}.
\end{equation}
As for the components $A_\mu^1$ and $A_\mu^2$, the situation is quite different. Because the ground state $\bm{\phi}_0$ can only provide mass to three of the four gauge fields, we need to set the massless gauge field to zero in the vortex solutions. To compute the mass terms related to the gauge field, we can insert the ground state $\bm{\phi}_0$ into the Lagrangian $\mathcal{L}_C$ and work out the terms containing $A_\mu^1$ and $A_\mu^2$ as
\be
\mathcal{L}_{A^2} & = & \frac{1}{4}(v_1^2+v_2^2)\left[ (A_\mu^0)^2+(A_\mu^1)^2+(A_\mu^2)^2+(A_\mu^3)^2\right] +\frac{1}{2}(v_1^2-v_2^2)A_\mu^0 A_\mu^3 \nonumber\\ 
& &{} +v_1 v_2A_\mu^0\left\{\cos[(n_1-n_2)\theta-\varphi] A_\mu^1-\sin[(n_1-n_2)\theta-\varphi] A_\mu^2 \right\}. 
\ee
The polar angle $\theta$ emerges. The mass eigenstates of the gauge field are only locally defined, conflicting with the central symmetry of vortex solutions. The simplest method to eliminate $\theta$ is to set $A_\mu^1$ and $A_\mu^2$ to zero everywhere. After doing so, the remaining $A_\mu$ becomes diagonal
\begin{equation}
	A_0(r\rightarrow \infty, \theta)=\begin{pmatrix}
		o(\frac{1}{r})	& 0 \\
		0	& o(\frac{1}{r})
	\end{pmatrix},\quad
	A_i(r\rightarrow \infty, \theta)=\begin{pmatrix}
		n_1\frac{\mathbf{e}_\theta}{r}+o(\frac{1}{r})	& 0 \\
		0	& n_2\frac{\mathbf{e}_\theta}{r}+o(\frac{1}{r})
	\end{pmatrix}
\end{equation}
where the remainder $o(\frac{1}{r})$ term decreases exponentially since $A_\mu^0,A_\mu^3$ are massive.

With non-zero $A_\mu^0,A_\mu^3$, the Lagrangian density becomes: 
\be
\mathcal{L}_C[\bm{\phi},A] & = &  \left|\partial_\mu\phi_1-i\frac{1}{2}(A_\mu^0+A_\mu^3)\phi_1\right|^2+\left|\partial_\mu\phi_2-i\frac{1}{2}(A_\mu^0-A_\mu^3)\phi_2\right|^2 \nonumber\\
& & {} -V_C(\bm{\phi}^\dagger\bm{\phi})+\frac{N_c}{4\pi}\epsilon^{\mu\nu\rho}\operatorname{Tr}\left( A_\mu\partial_\nu A_\rho-i\frac{2}{3} A_\mu A_\nu A_\rho\right),
\ee
In this special case where we have two scalar fields, $A_\mu^+=(A_\mu^0+ A_\mu^3)/2$ couples to $\phi_1$ and propagates unit quark number, while $A_\mu^-=(A_\mu^0- A_\mu^3)/2$ couples to $\phi_2$ and also propagates unit quark number. This two-flavor system behaves much like two copies of the one-flavor case, which simplifies the calculations and makes it more convenient to work with
\begin{equation}
	A_\mu=\begin{pmatrix}
		A^+_\mu	& 0 \\
		0	& A^-_\mu
	\end{pmatrix},\quad
		A_\mu^{+,-}(r\rightarrow \infty, \theta)=\left(o\!\left(\frac{1}{r}\right),n_{1,2}\frac{\mathbf{e}_\theta}{r}+ o\!\left(\frac{1}{r}\right)\right).
	\end{equation}

The non-Abelian Chern-Simons term also induces a current
\be
J^{\mu,a} & = & \frac{N_c}{4\pi}\epsilon^{\mu\nu\rho}\partial_\nu A_\rho^a =\frac{N_c}{8\pi}\epsilon^{\mu\nu\rho} F_{\nu\rho}^a, \; a=0 , 1, 2, 3,
\ee
with $F_{\mu\nu}$ being the field strength tensor of $A_\mu$. Actually, since the vortexes are with central symmetry which implies $[A_\mu,A_\nu]=0$, the currents for $A_\mu^\pm$ become
\be
J^{\mu,\pm} & = & J^{\mu,0}\pm J^{\mu,3}=\frac{N_c}{2\pi}\epsilon^{\mu\nu\rho}\partial_\nu A_\rho^\pm.
\ee
Therefore, we can calculate the flux of the vortexes:
\begin{equation}
	\Phi^+=\int \epsilon^{0\nu\rho}\partial_\nu A_\rho^+ dxdy=2\pi n_1,\quad \Phi^-=\int \epsilon^{0\nu\rho}\partial_\nu A_\rho^- dxdy=2\pi n_2,
\end{equation}
and the topological charge
\begin{equation}
	Q^+=\int J^{0,+} dxdy=n_1 N_c, \quad 	Q^-=\int J^{0,-} dxdy=n_2 N_c.
\end{equation}

If two vortexes are located at a considerable distance from each other and interact very weakly, the flux and charges associated with different gauge fields do not influence each other. This implies that the total spin induced by the different flux and topological charges can be simply added up:
\begin{equation}
	S=\frac{\Phi^+ Q^+}{4\pi}+\frac{\Phi^- Q^-}{4\pi}=(n_1^2+n_2^2) \frac{N_c}{2}, \; n_1,n_2\in\mathbb{Z},
\end{equation}
and the baryon number can be defined as the one-flavor case
\begin{equation}
	B=\frac{Q^+ +Q^-}{N_c}=n_1+n_2.
\end{equation}
As is obvious, the baryon number is the sum of the winding numbers of all flavors and, the winding number of each flavor contributes independently. For example, choosing $n_1=1, n_2=0$ results in a situation that is similar to that of a one-flavor scenario, this does not imply that the second flavor quark does not function since $\phi_2\neq 0$ enters the vortex solution. In the case where $n_1=n_2=1$, the vortex would contain a doubled baryon number and a spin $N_c$. This structure is commonly referred to as a two-baryon system. On the other hand, if $n_1=1$ and $n_2=-1$, the vortex could be understood as a baryon-antibaryon structure, with zero baryon number and a spin of $N_c$. Therefore, every baryon or anti-baryon within a vortex on average carries at least $N_c/2$ spin. As a consequence, vortex solutions that contain multiple flavors cannot avoid having high spins. It has also been suggested that the description of high-spin baryons should be based on Hall droplets or vortexes. And normal low-spin baryons should utilize the Skyrme model~\cite{Komargodski:2018odf}. It is also worth noting that only a particular type of vortex solution is taken into account in $\mathcal{L}_C$. Other possible vortex configurations are more complex and may exhibit different behavior, which deserves investigation in future research.

   While finding all vortex solutions is challenging in multi-flavor situations, an alternative approach involves investigating the baryon structure on two-dimensional objects, In references 
   \cite{Eto:2023tuu,Eto:2023rzd}, The authors have discovered that, owing to the anomalous coupling of the $\eta^\prime$ meson to rotation \cite{Huang:2017pqe,Huang:2019rkz,Nishimura:2020odq}, the ground state of rapidly rotating two-flavor QCD matter forms a chiral soliton lattice (CSL),which can be either Abelian or non-Abelian \cite{Eto:2021gyy}. Additionally, there exists a domain-wall skyrmion phase located on the CSL. In the case of non-Abelian CSL, the domain walls are referred to as up and down solitons, both composed of $\eta^\prime$ and $\pi 0$ meson fields. Interestingly, the common 3+1 dimensional Skyrmions flow into the 3+1 dimensional domain walls, transforming into baby Skyrmions. This unified viewpoint connects baryon dynamics across different dimensions \cite{Nitta:2012wi,Eto:2015uqa,Gudnason:2014nba,Nitta:2022ahj}. While our vortex approach appears more effective for the one-flavor case, the baby skyrmion picture proves more powerful for scenarios involving multiple flavors. We believe that a deeper connection exists between these two approaches and they should be combined to provide a complete description for arbitrary flavors. Additionally, it is worth mentioning that under a magnetic field, domain-wall Skyrmion chains also emerge in the chiral soliton lattice \cite{Eto:2023lyo,Eto:2023wul,Qiu:2024zpg,Chen:2021vou,Fukushima:2018ohd,Son:2007ny}.

\section{Discussion and Conclusion}

In the large $N_c$ limit, we explore a concrete theory living on the $\eta'$ domain wall. By using level-rank duality, the global flavor symmetry is gauged, and the fermionic matter field is bosonized. For the one-flavor case, this theory is identified with the Chern-Simons-Higgs theory, which features a series of topological vortex solutions. The vortex solution with one winding number carries one unit topological charge and induces a spin of $N_c/2$, which could reasonably be interpreted as a baryon in the 2+1 dimensional spacetime. A more detailed analysis reveals that these vortex solutions share the same large $N_c$ properties as the baryons in terms of mass, radius, and interaction behavior. 
However, in the leading order of the $N_c$ expansion, the kinetic term of the gauge field does not emerge. This suggests that the self-interaction of gluons is suppressed on the domain wall. Through the particle-vortex duality, the Chern-Simons-Higgs theory is dual to the Zhang-Hansson-Kivelson theory, which is developed to describe the fractional Hall effect. This supports the conjecture that baryons in 2+1 dimensions can be represented as Hall droplets, and also shows that quarks obey fractional statistics.

Generalizing to more flavors becomes more complicated because the global flavor symmetry group includes the baryon number symmetry and the isospin symmetry. After the latter is gauged, the gauge fields become entangled, making it difficult to find vortex solutions. For the specific choice we have adopted, the vortex solutions for multiple flavors simply behave as a superposition of many one-flavor vortex solutions. Nonetheless, the multi-flavor vortexes all carry high spin, which entails that only high-spin baryons can be described in this framework. The lower-spin baryons should be considered as Skyrmions, which are soliton structures of pion fields, without $\eta'$ field.To establish a connection between the common skyrmion in 3+1 dimensions and the baryon on the domain wall for the multi-flavor case, the description of baby skyrmions is likely convenient. However, the direct relationship between baby skyrmions and our vortex remains unclear \cite{Eto:2023tuu,Eto:2023rzd,Huang:2017pqe,Huang:2019rkz,Nishimura:2020odq,Eto:2021gyy,Nitta:2012wi,Eto:2015uqa,Gudnason:2014nba,Nitta:2022ahj,Eto:2023lyo,Eto:2023wul,Qiu:2024zpg,Chen:2021vou,Fukushima:2018ohd,Son:2007ny}.

 Our proposal that vortex solutions can be associated with baryons in 2+1 dimensions significantly hinges on the dynamics of the $\eta'$ field. In the dense baryonic matter, it is possible that that baryons are fractional quantum Hall droplets, or vortexes~\cite{Ma:2019ery,Rho:2022meo,Ma:2020nih} and it is indeed found skyrmion crystal approach dense nuclear matter transit to a layer structure~\cite{Park:2019bmi,Harada:2016tkf,Kawaguchi:2018fpi,Ma:2021nuf,Brown:2010api}. In addition, our work might also provide a bridge across two fields, condensed matter theory and high energy physics, worthy of future investigation.

\section*{Acknowledgements} 

The work of Y.~L. M. is supported in part by the National Science Foundation of China (NSFC) under Grant No. 12347103, No. 11875147 and No. 12147103 and the National Key R\&D Program of China under Grant No. 2021YFC2202900.



\end{document}